\documentclass[a4paper]{article}

\usepackage{amsmath,amsfonts,amssymb}
\usepackage{amsthm}
\usepackage{boxedminipage}
\usepackage{framed}
\usepackage{thmtools}
\usepackage{a4wide}

\usepackage{dsfont}
\usepackage{xifthen}
\usepackage{tabularx}
\usepackage{tikz}
\usepackage[numbers]{natbib}
\usepackage[hidelinks]{hyperref}
\usepackage[capitalize,noabbrev]{cleveref}

\newtheorem{theorem}{Theorem}
\newtheorem{proposition}{Proposition}

\usetikzlibrary{positioning,backgrounds,patterns,calc,matrix,shapes,decorations.pathreplacing,decorations.pathmorphing,math}

\tikzset{vertex/.style ={circle,inner sep=0pt,minimum size=4pt}}
\tikzstyle{path} = [color=black,opacity=.15,line cap=round, line join=round, line width=6pt]

\DeclareMathOperator{\dist}{dist}

\newlength{\RoundedBoxWidth}
\newsavebox{\GrayRoundedBox}
\newenvironment{GrayBox}[1]%
   {\setlength{\RoundedBoxWidth}{.95\textwidth}
    \def\boxheading{#1}
    \begin{lrbox}{\GrayRoundedBox}
       \begin{minipage}{\RoundedBoxWidth}}%
   {   \end{minipage}
    \end{lrbox}
    \begin{center}
    \begin{tikzpicture}%
       \node(Text)[draw=black!20,fill=white,rounded corners,%
             inner sep=2ex,text width=\RoundedBoxWidth]%
             {\usebox{\GrayRoundedBox}};
        \coordinate(x) at (current bounding box.north west);
        \node [draw=white,rectangle,inner sep=3pt,anchor=north west,fill=white] 
        at ($(x)+(6pt,.75em)$) {\boxheading};
    \end{tikzpicture}
    \end{center}}     

\newenvironment{defproblemx}[2][]{\noindent\ignorespaces%
                                \FrameSep=6pt%
                                \parindent=0pt%
                \vspace*{-1.5em}
                \ifthenelse{\isempty{#1}}{%
                  \begin{GrayBox}{\textsc{#2}}%
                }{%
                  \begin{GrayBox}{\textsc{#2}  parameterized by~{#1}}%
                }
                \begin{tabular*}{\textwidth}{@{\hspace{.1em}} >{\itshape} p{1.5cm} p{0.86\textwidth} @{}}
            }{
                \end{tabular*}%
                \end{GrayBox}%
                \ignorespacesafterend
            }

\newcommand{\defproblema}[3]{
  \begin{defproblemx}{#1}
    Input:  & #2 \\
    Question: & #3
  \end{defproblemx}
}%

\newcommand{\vte}{\textsc{Variable-Length Traffic Engineering with Waypoints}}
\newcommand{\fste}{\textsc{Fixed-Length Splittable Traffic Engineering with Waypoints}}

\date{}

\begin{document}

\title{The Complexity of Traffic Engineering in Segment Routing Networks}

\author{Matthias Bentert \and Stefan Schmid}

\maketitle

\begin{abstract}
Segment routing is a modern variant of source routing which has recently become popular in the Internet.
It allows operators to equip network packets with an additional header which specifies segments: a list of \emph{waypoints} the packet needs to traverse on its way to the destination.
This is reminiscent of Valiant routing in parallel computing and introduces an additional degree of freedom in traffic engineering.
This flexibility can be exploited to minimize congestion in the network. 

This paper sheds light on the computational complexity of traffic engineering in segment routing networks. We consider two problem variants: a fixed-length variant where link weights are given and only waypoints can be chosen, and a variable-length variant where additionally the network operator can also choose the link weights, which then define the shortest paths taken between the waypoints.
We prove that the first variant is NP-hard even if the network graph is planar and if all demands and link weights are 1.
We also show that the second variant is NP-hard already for planar network graphs with all demands being 1.
On the positive side, we present a compact and general mixed integer linear program to compute exact solutions for the first variant.
\end{abstract}

\section{Introduction}

To make best use of their infrastructure and to provide good performance to their customers, Internet Service Providers~(ISPs) employ traffic engineering~(TE) to select paths for their communication traffic that minimize congestion. 
Segment routing~(SR) has recently emerged as a powerful approach to perform~TE, which received much attention in industry and standardization bodies (the IETF). 
Traditionally, packets are restricted to follow shortest paths which are indirectly defined by the link weights which can be set by network operators. In SR networks, there is an additional degree of freedom: packets can be equipped with a packet header defining a list of 
\emph{waypoints}: nodes (essentially IPv6 addresses) which the packet needs to traverse on its way to the destination.\footnote{Technically and more generally, segments are instructions (e.g., forwarding to an interface) that are executed on subsequent nodes in the network.} Between each pair of waypoints as well as from the source to the first waypoint and from the last waypoint to the destination (called \emph{segments}), a packet takes a shortest path, defined as usual, based on link weights.
If there are multiple shortest paths, traffic is split across them on a per-flow granularity (to minimize packet reordering).
Segment routing can hence be seen as a modern variant of source routing and is inspired by Valiant routing known in parallel computing~\cite{valiant1982scheme}. 
Empirical studies show that such a redirection, already with a single waypoint, can significantly improve TE performance in practice~\cite{SR-Thesis-Aubry,SR-survey21}. 

Segment routing networks introduce a novel graph-theoretical optimization problem: 
how to select waypoints in order to minimize congestion? 
This problem is interesting already in scenarios where link weights are \emph{given}, and we will refer to this problem version as the \emph{fixed-length} variant in the following (interpreting weights as distances). Here, we also consider settings which support \emph{splitting ratios}~\cite{MChiesaetal}: scenarios where network operators can decide to assign different amounts of traffic to  different shortest paths.
The problem version where in addition to the waypoints, also the link weights can be chosen, is referred to as the \emph{variable-length} variant.

This paper is concerned with these algorithmic problems underlying segment routing. 
Specifically, we investigate the case of a single waypoint per demand.
Given the high interest of industry in the topic, surprisingly little is known today about the graph-theoretical challenges of traffic engineering with waypoints. Parham et al.~\cite{parham2021traffic} showed that a \emph{joint} optimization of link weights and waypoints can in principle significantly improve traffic engineering compared to scenarios where link weights and waypoints are optimized separately. 
Amiri et al.~\cite{amiri2018waypoint,amiri2018charting,akhoondian2020walking} initiated the study of the parametrized complexity of routing through waypoints in a related problem in which the task is to find a shortest walk visiting a given set of waypoints and respecting edge capacities; they presented exact polynomial-time algorithms for graphs of bounded treewidth or restricted problems where the number of waypoints is logarithmically bounded.
Recently, Bazgan et al.~\cite{bazgan2025parameterized} showed that the fixed-length segment routing problem is NP-hard even for graphs with constant treewidth and when only one waypoint per demand is allowed. In our work, we complement these results, among others (see below), by showing that both the fixed-length variant (for which we also consider splitting ratios which Bazgan et al.\ did not) and the variable-length variant are NP-hard even on the practically relevant class of planar graphs (many ISP topologies are planar~\cite{pignolet2017tomographic}).

\subsection{Additional Related Work}

Traffic engineering is a crucial concern of ISPs and an evergreen topic in networking~\cite{chiesa2016lying,Fortz2004,kumar2018semi,racke2019compact,zuzic2023sparse}.
Many innovations in networking over the last years were at least partially motivated by the desire to improve TE~\cite{feamster2014road,parham2021traffic}.
There already exists a large body of literature 
on TE, for various network protocols,
from MPLS~\cite{gupta2005traveling,mpls-te} over SDN~\cite{sdn-te}
to SR~\cite{ietf-SR-arch}. 
There are also already several well-known analytical and complexity results on the
achievable performance of TE in traditional networks~\cite{MChiesaetal,Fortz2004,MaxECMP2002}.

Segment routing can be used on top of MPLS networks and on IPv6 networks~\cite{ietf-SR-arch}. 
Many more applied results on traffic engineering for 
segment routing are already available~\mbox{\cite{SR-Thesis-Aubry,bhatia2015,SR-Thesis-Renaud,moreno2017traffic}} and we refer the interested reader to a recent survey by~\citet{SR-survey21}.
Moreno et al.~\cite{moreno2017traffic} showed that
a minimal number of stacked labels suffice to exploit the benefits of segment routing successfully. Aubry et al.~\cite{SR-Thesis-Aubry,aubry2018robustly} lay the algorithmic foundations of segment routing, considering different and related applications.
\citet{bhatia2015} consider the problem
of determining the optimal
parameters for segment routing in offline and online scenarios;
primarily focusing on 2-segment routing where traffic between the ingress and egress passes
through exactly one intermediate node.  
They develop a traffic-matrix oblivious algorithm for robust segment routing in the offline setting. 
De Boeck~\cite{de2023case} considers stochastic approaches to deal with uncertainty in segment routing traffic engineering. 
There also exists algorithmic work on local fast-rerouting for SR networks, as a reaction to link failures~\cite{DBLP:conf/infocom/FoersterPCS18}.

\subsection{Our Contribution}

Our main contributions concern the computational complexity of traffic engineering in segment routing networks.
We first rigorously prove that even the fixed-length problem variant (where link weights are given and traffic can be distributed across paths according to operator-defined splitting ratios) is NP-hard already if the network graph is planar and if all demands and link weights are 1.
We further show that the variable-length problem variant is also NP-hard already for planar network graphs, even with all demands being 1.
Complementing these negative results, we present a compact and general mixed integer linear program to compute exact solutions.
We next present the high-level idea of our two hardness results.

For the fixed-length variant, we reduce from multicolored clique. At a high level, we construct horizontal and vertical canonical paths representing the vertices of the input graph, grouped into bundles corresponding to colors. A feasible routing is forced to select one canonical path from each bundle. Where a horizontal and a vertical path meet, a small gadget determines whether both can be used simultaneously without exceeding the congestion bound: this is possible precisely if the represented vertices share an edge (or are the same vertex when considering two vertices of the same color).

For the variable-length variant, we instead reduce from a planar SAT variant. The weight assignment effectively encodes a truth assignment: in each variable gadget, making one of two alternative paths shorter forces the corresponding unique shortest paths to represent either true or false. Clause gadgets are connected to these choices such that their terminal pairs can be routed without congestion only through a literal consistent with the variable assignment. Thus, a congestion-one routing exists exactly when all clauses can simultaneously be satisfied.

\section{Preliminaries and Formal Model}\label{sec:prelim}

For a positive integer~$n$, we denote by~$[n]$ the set~$\{1,2,\ldots,n\}$.
We denote the number of nodes (e.g., routers) in a graph by~$n$ and the number of links by~$m$.
Let~$G = (V,E)$ be a (directed or undirected) graph. Moreover, let~$T = \{(s_1,t_1,r_1),(s_2,t_2,r_2),\ldots,(s_k,t_k,r_k)\}$ be a set of triples where~$s_i,t_i \in V$ are nodes (called terminals) and~$r_i \in \mathds{Q}^+$ is a positive number.
Let~$w \colon E \rightarrow \mathds{Q}^+$ be a link-weight function.
The task is to select one node~$p_i$ (called a waypoint) and an~$r_i$-flow between~$s_i$ and~$p_i$ and between~$p_i$ and~$t_i$ for each triple~$(s_i,t_i,r_i) \in T$ such that each path of the flow is a shortest path (with respect to~$w$) between the two respective endpoints ($p_i$ and the terminals~$s_i$/$t_i$).
The goal is to minimize the maximum flow over any link in the graph.
The load of a link~$e$ is~$\sum_{P \in P_e} r_P$, where~$P_e$ is the set of paths in the solution that contain the link~$e$ and~$r_P$ is the flow sent over the path~$P$.

We study the following two problem variants.
Note that in one of the variants, the link-weight function~$w$ is given as input, whereas in the second, we are free to define the function as part of the solution.
In the first variant, flows can be split according to an optimizable ratio. In the second variant, flows take unique paths.

\defproblema{\fste}
{A graph~$G=(V,E)$, a set ${T = \{(s_1,t_1,r_1),\ldots,(s_k,t_k,r_k)\}}$ of triples where~$s_i, t_i \in V$ and~$r_i \in \mathds{Q}^+$ for all~$i \in [k]$, a link-weight function~$w \colon E \rightarrow \mathds{Q}^+$, and a positive number~$\ell \in \mathds{Q}^+$.}
{Is there a tuple~$Q=(p_1,p_2,\ldots,p_k)$ of nodes and for each~$i \in [k]$ two $r_i$-flows~$F^1_i$ and~$F^2_i$ such that~$F^1_i$ is a collection of shortest~$s_i$-$p_i$-paths, $F^2_i$ is a collection of shortest~$p_i$-$t_i$-paths, and at most~$\ell$ units of flow are sent over each link in the graph?}

\defproblema{\vte}
{A graph~$G=(V,E)$, a set ${T = \{(s_1,t_1,r_1),\ldots,(s_k,t_k,r_k)\}}$ of triples where~$s_i, t_i \in V$ and~$r_i \in \mathds{Q}^+$ for all~$i \in [k]$, and a positive number~$\ell \in \mathds{Q}^+$.}{Does a tuple~$Q=(p_1,p_2,\ldots,p_k)$ of nodes and a link-weight function~$w \colon E \rightarrow \mathds{Q}^+$ exist such that for each~$i \in [k]$, there is a unique shortest path~$P^1_i$ between~$s_i$ and~$p_i$ and a unique shortest path~$P^2_i$ between~$p_i$ and~$t_i$ and ${\max\limits_{e \in E} \sum\limits_{i \in [k]} \sum\limits_{P \in \{P^1_i,P^2_i\}} e_P \cdot r_i \leq \ell}$? Here, $e_P$ is~1 if~$e$ is a link in~$P$ and 0 otherwise.}

\section{Analysis and Results}\label{sec:analysis}

In this section, we show our main results. We start by proving that the fixed-length variant is NP-hard even in planar graphs with unit demands.

\begin{theorem}
    \label{thm:fste}
    \fste{} is NP-hard even if the input contains an undirected planar graph,~$\ell = 1$, $r_i=1$ for all~$i \in [k]$, and~${w(e) = 1}$ for all~$e \in E$.
\end{theorem}

\begin{proof}
    We show a reduction from \textsc{Multicolored Clique}.
    Therein, one is given an undirected graph~${G=(V,E)}$, an integer~$k$, and a coloring function~$c\colon V \rightarrow [k]$.
    The question is whether there is a multicolored clique in~$G$, that is, a set of pairwise adjacent nodes containing exactly one node of each color in~$[k]$.
    It is known that \textsc{Multicolored Clique} remains NP-hard even when~$k \geq 4$ and there are the same number~$\nu = \frac{n}{k}$ of nodes of each color~\cite{GJ79}.
    We provide a polynomial-time reduction from such an instance~$(G,c,k)$ of \textsc{Multicolored Clique} to an instance~$(G' = (V',E'),T,w,1)$ of \fste, where~$w(e) = 1$ for all~$e \in E'$.
    We mention that the construction is similar and based on a reduction by \mbox{\citet{BNRZ23}}.

    Before we give the overall construction, we start with four simple gadgets, which we call \emph{pass-through}, \emph{intersection}, \emph{blocking}, and \emph{shortcut} gadgets, respectively.
    They are depicted in \cref{fig:gadgets} and the intuitive ideas for them are as follows.
    
    \begin{figure*}[t]
        \centering
        \begin{tikzpicture}
            \def\x{3}
            \node[circle,draw] at(-\x,0) (c) {};
            \node[circle,draw] at(-\x,1) (n) {} edge(c) edge(-\x,1.5);
            \node[circle,draw] at(-\x-1,0) (w) {} edge(c) edge(-\x-1.5,0); 
            \node[circle,draw] at(-\x+1,0) (e) {} edge(c) edge(-\x+1.5,0); 
            \node[circle,draw] at(-\x,-1) (s) {} edge(c) edge(-\x,-1.5); 

            \node[circle,draw] at(-.5,.5) (st) {} edge(-1,.5) edge(-.5,1);
            \node[circle,draw] at(0,0) (m) {} edge(st);
            \node[circle,draw] at(.5,-.5) (ta) {} edge(m) edge(1,-.5) edge(.5,-1);

            \foreach \i in {1,2,3,4}{
                \node[circle,draw,inner sep=2pt] at(\x-1,1.25-.5*\i) (x1\i) {} edge(\x-1.5,1.25-.5*\i);
                \node[circle,draw,inner sep=2pt] at(\x-.5,1.25-.5*\i) (x2\i) {} edge(x1\i);
                \node[circle,draw,inner sep=2pt] at(\x,1.25-.5*\i) (x3\i) {} edge(x2\i);
                \node[circle,draw,inner sep=2pt] at(\x+.5,1.25-.5*\i) (x4\i) {} edge[very thick, red](x3\i);
                \node[circle,draw,inner sep=2pt] at(\x+1,1.25-.5*\i) (x5\i) {} edge(x4\i);
                \node[circle,draw,inner sep=2pt] at(\x+1.5,1.25-.5*\i) (x6\i) {} edge(x5\i) edge(\x+2,1.25-.5*\i);
            }
            \foreach \i in {1,2,3}{
				\pgfmathtruncatemacro\ii{\i+1};
                \foreach \j in {1,2,...,6}{
                    \draw (x\j\i) -- (x\j\ii);
                }
            }
            \node[circle,draw,label=$s_1$,inner sep=2pt] (s1) at(\x-1,1.25) {} edge(x11);
            \node[circle,draw,label=$s_2$,inner sep=2pt] (s2) at(\x-.5,1.25) {} edge(x21);
            \node[circle,draw,label=$s_3$,inner sep=2pt] (s3) at(\x,1.25) {} edge(x31);
            \node[circle,draw,label=$t_3$,inner sep=2pt] (t3) at(\x+.5,1.25) {} edge(x41);
            \node[circle,draw,label=$t_2$,inner sep=2pt] (t2) at(\x+1,1.25) {} edge(x51);
            \node[circle,draw,label=$t_1$,inner sep=2pt] (t1) at(\x+1.5,1.25) {} edge(x61);

    		\begin{pgfonlayer}{background}
                    \draw[path] (\x-1.5,-.25) -- (\x+2,-.25);
                    \draw[path] (s1.center) -- (x14.center) -- (x64.center) -- (t1.center);
                    \draw[path] (s2.center) -- (x22.center) -- (x52.center) -- (t2.center);
                    \draw[path] (s3.center) -- (x31.center) -- (x41.center) -- (t3.center);
            \end{pgfonlayer}       

            \node[circle,draw,label=right:$t_1$] (y1) at(2*\x,1) {} edge(2*\x,1.5);
            \node[circle,draw,label=right:$s_1$] (y2) at(2*\x,.333) {} edge(y1);
            \node[circle,draw,label=right:$s_2$] (y3) at(2*\x,-.333) {} edge[dashed](y2);
            \node[circle,draw,label=right:$t_2$] (y4) at(2*\x,-1) {} edge(y3) edge(2*\x,-1.5);
        \end{tikzpicture}
        \caption{The four gadgets used in the proof of \cref{thm:fste}. The leftmost one is a pass-through gadget, the second is an intersection gadget, the third is a blocking gadget, and the last one is a shortcut gadget (where the dashed edge represents a path of length~$25n$).}
        \label{fig:gadgets}
    \end{figure*}
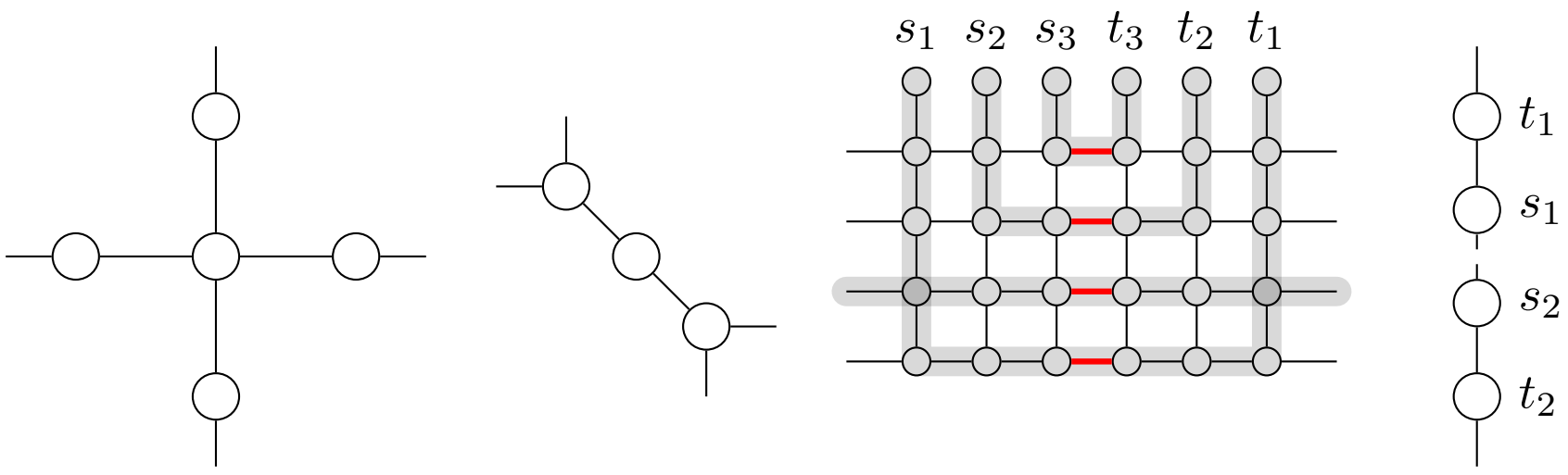
    For the pass-through gadget and the intersection gadget, consider the case that a path from the top goes to the bottom and another path comes from the left and goes to the right.
    In the pass-through gadget, these paths can be routed so as to avoid any overlap in the links used.
    This is not possible in the intersection gadget.
    For the blocking gadget, we have some number~$q$ of paths going from left to right and~$q-1$ terminal pairs.
    The maximum flow that can be routed without assigning a flow of more than one to some link is~$q$ as the~$q$ red links cut the left from the right.
    Since there is a flow of~$q-1$ coming from the~$q-1$ terminal pairs, at most one additional unit of flow can be routed through the gadget.
    Finally, the shortcut gadget allows us to have relatively short paths between the terminals while forcing the solution to include much longer paths.
    Consider a shortcut gadget as depicted on the right side of \cref{fig:gadgets} (with the same node names).
    We show that we can assume without loss of generality that in any solution for~$\ell=1$ a flow of one is sent directly over the link~$\{s_1,t_1\}$ for the triple~$(s_1,t_1,1)$ and for the triple~$(s_2,t_2,1)$, a flow of one is sent over the link~$\{s_2,t_2\}$.
    To this end, let~$q$ be the flow that is sent over the link~$\{s_1,t_1\}$ for the triple~$(s_1,t_1,1)$ and let~$r$ be the amount of flow sent over the link~$\{s_2,t_2\}$ for the triple~$(s_2,t_2,1)$.
    Note that for the triple~$(s_1,t_1,1)$ a flow of~$1-q$ is sent over the path from~$s_1$ to~$s_2$ and over the link~$\{s_2,t_2\}$.
    Analogously for the triple~$(s_2,t_2,1)$, a flow of~$1-r$ is sent over the link~$\{s_1,t_1\}$ and the~$s_1$-$s_2$-path.
    Since we send at most one unit of flow over any link in a solution for~$\ell=1$, it holds that~$q + (1-r) \leq 1$ (because of link~$\{s_1,t_1\}$) and~$(1-q) + r \leq 1$ (because of link~$\{s_2,t_2\}$).
    This implies that~$q=r$ and hence we have that exactly one unit of flow is sent over these two links for the two triples~$(s_1,t_1,1)$ and~$(s_2,t_2,1)$.
    Hence, there is no capacity to send any flow for other triples over these links and since setting~$q=1=r$ is a solution that sends one unit flow over these two links and no flow over any other links, we can assume without loss of generality that this is part of an optimal solution.
    For technical reasons, we then subdivide each link in each blocking gadget four times.
    
    We continue with the complete construction, which is illustrated in \cref{fig:hardness-example}.
	\begin{figure*}[t]
		\begin{tikzpicture}[xscale=.8,yscale=.8]
    		\newcommand{\linkInKDSPInstance}{
    			\node[vertex,fill=pink,minimum size=5pt] at (\i + \j * \smallDist - 2 * \smallDist,2 * \smallDist + -\x - \y * \smallDist) {};
                \node[vertex,fill=pink,minimum size=5pt] at (\x + \y * \smallDist - 2 * \smallDist,2 * \smallDist + -\i - \j * \smallDist) {};
    		}
            \newcommand{\colorA}{blue}
			\newcommand{\colorB}{red!90!black}
			\newcommand{\colorC}{green!60!black}
			\newcommand{\colorD}{black}
			
			\newcommand{\smallDist}{0.2}
			\newcommand{\smallerDist}{0.1}
			\newcommand{\cliqueVertices}{1, 3, 2, 1} 
		
			\begin{scope}[yshift=-1cm]
				\foreach \x in {1,...,3}{
					\node[vertex,fill=\colorA,
						label=above:{$1,\x$}
						] (v-1\x) at (\x,0) {};
					\node[vertex,fill=\colorB,
						label=right:{$2,\x$}
						] (v-2\x) at (4,-\x) {};
					\node[vertex,fill=\colorC,
						label=below:{$3,\x$}
						] (v-3\x) at (\x,-4) {};
					\node[vertex,fill=\colorD,
						label=left:{$4,\x$}
						] (v-4\x) at (0,-\x) {};
				}
			\end{scope}
			
			\begin{scope}[xshift=6cm, xscale=1.4, yscale=1.2]
			
				\foreach[count=\i] \color in {\colorA,\colorB,\colorC,\colorD}{
					\node[vertex,fill=\color,label=below:{$s_\i$}] (s-\i) at (0,-\i) {};
					\node[vertex,fill=\color,label=below:{$t_\i$}] (t-\i) at (5,-\i) {};
					
					\pgfmathtruncatemacro\j{\i + 4};
					
					\node[vertex,fill=\color,label=right:{$t_\j$}] (tt-\i) at (\i,-5) {};
					\node[vertex,fill=\color,label=right:{$s_\j$}] (ss-\i) at (\i,0) {};
					\foreach \j in {1,...,3}
					{
						\node[vertex,fill=\color] (sc-\j\i) at (\smallDist,2 * \smallDist + -\i - \j * \smallDist) {};
						\draw[dotted,thick] (sc-\j\i) -- (s-\i);
						\node[vertex,fill=\color] (tc-\j\i) at (4.7,2 * \smallDist + -\i - \j * \smallDist) {};
						\draw[dotted,thick] (tc-\j\i) -- (t-\i);
						\draw (tc-\j\i) -- (sc-\j\i);

						\node[vertex,fill=\color] (ssc-\j\i) at (\i + \j * \smallDist - 2 * \smallDist,-\smallDist) {};
						\draw[dotted,thick] (ssc-\j\i) -- (ss-\i);
						\node[vertex,fill=\color] (ttc-\j\i) at (\i + \j * \smallDist - 2 * \smallDist,-4.7) {};
						\draw[dotted,thick] (ttc-\j\i) -- (tt-\i);
						\draw (ttc-\j\i) -- (ssc-\j\i);
					}
				}
                \foreach \i in {0,1,2,3,4}{
                    \foreach \j in {1,2,3,4}{
                        \draw[very thick,violet] (\i+.4,.4-\j) -- (\i+.4,-.3-\j);
                        \draw[very thick,violet] (\i+.4,-.3-\j) -- (\i+.6,-.3-\j);
                        \draw[very thick,violet] (\i+.6,-.3-\j) -- (\i+.6,.4-\j);
        				
                        \draw[very thick,violet] (-.4+\j,-.4-\i) -- (\j+.3,-.4-\i);
                        \draw[very thick,violet] (\j+.3,-.4-\i) -- (\j+.3,-.6-\i);
                        \draw[very thick,violet] (\j+.3,-.6-\i) -- (-.4+\j,-.6-\i);
                    }
                }
                			
					\foreach \i in {1,...,4} \foreach \x in {1,...,4} \foreach \j in {1,...,3} \foreach \y in {1,...,3}
					{
						\node[vertex,fill=orange] at (\i + \j * \smallDist - 2 * \smallDist,2 * \smallDist + -\x - \y * \smallDist) {};
						\node[vertex,fill=orange] at (\x + \y * \smallDist - 2 * \smallDist,2 * \smallDist + -\i - \j * \smallDist) {};
					}

					\foreach \i in {1,...,4} \foreach \j in {1,...,3} \foreach \y in {1,...,3}
					{
						\ifnum \j=\y%
							\pgfmathtruncatemacro\x{\i};
							\linkInKDSPInstance
						\fi%
					}

					\foreach \i / \j / \x / \y in {
							1/3/3/3, 1/2/3/3, 3/1/4/3, 3/3/2/1, 1/3/4/1, 2/2/1/3, 1/1/4/3, 3/3/4/2, 1/2/2/2, 1/2/3/1, 2/1/4/3, 2/2/4/2, 2/3/3/1, 4/2/3/2
						} 
					{
						\draw (v-\i\j) -- (v-\x\y);
						\linkInKDSPInstance
					}
                   \node[vertex,fill,label=$c_1$] (c1) at(-.5,.5) {} edge[dashed](s-1) edge[dashed](s-2) edge[dashed](s-3) edge[dashed](s-4) edge[dashed](ss-1) edge[dashed](ss-2) edge[dashed](ss-3) edge[dashed](ss-4);
                   \node[vertex,fill,label=above left:$c_2$] (c2) at(5.5,-5.5) {} edge[dashed](t-1) edge[dashed](t-2) edge[dashed](t-3) edge[dashed](t-4) edge[dashed](tt-1) edge[dashed](tt-2) edge[dashed](tt-3) edge[dashed](tt-4);
                   \draw[smooth] (c1) -- (5.5,.5) -- (c2);
                    \foreach[count=\i] \j in \cliqueVertices
					{
						\foreach[count=\x] \y in \cliqueVertices
						{
							\ifnum \i<\x%
								\draw[very thick] (v-\i\j) -- (v-\x\y);
								\linkInKDSPInstance
							\fi
						}
					\begin{pgfonlayer}{background}
						\draw[path] (s-\i.center) -- (sc-\j\i.center) -- (tc-\j\i.center) -- (t-\i.center);
						\draw[path] (ss-\i.center) -- (ssc-\j\i.center) -- (ttc-\j\i.center) -- (tt-\i.center);
				\end{pgfonlayer}
					}
			\end{scope}
			
		\end{tikzpicture}
		\caption{
			An illustration of the reduction from \textsc{Multicolored Clique} to \fste. \\
			\emph{Left side:} Example instance for \textsc{Multicolored Clique} with~$k=4$ colors and $\nu = 3$ nodes per color.
            Each node is given a label~$(i,j)$, where~$i$ indicates the color of a vertex and~$j$ is an index of the respective node within each color class.
			A multicolored clique is highlighted (by thick links). \\
			\emph{Right side:} The constructed instance.
            The pink nodes represent pass-through gadgets, the orange nodes represent intersection gadgets, the violet objects represent blocking gadgets, and dashed links indicate shortcut gadgets.
            Dotted links (at the beginning and end of each path) indicate paths of length~$25n = 300$.
            The eight solution paths corresponding to the solution on the left are highlighted (not including the solution paths for terminals in blocking or shortcut gadgets).
            The waypoint can be any node on the highlighted path close enough to the middle.
			Note that all highlighted solution paths intersect in pass-through gadgets and hence are link-disjoint.
		}
		\label{fig:hardness-example}
	\end{figure*}
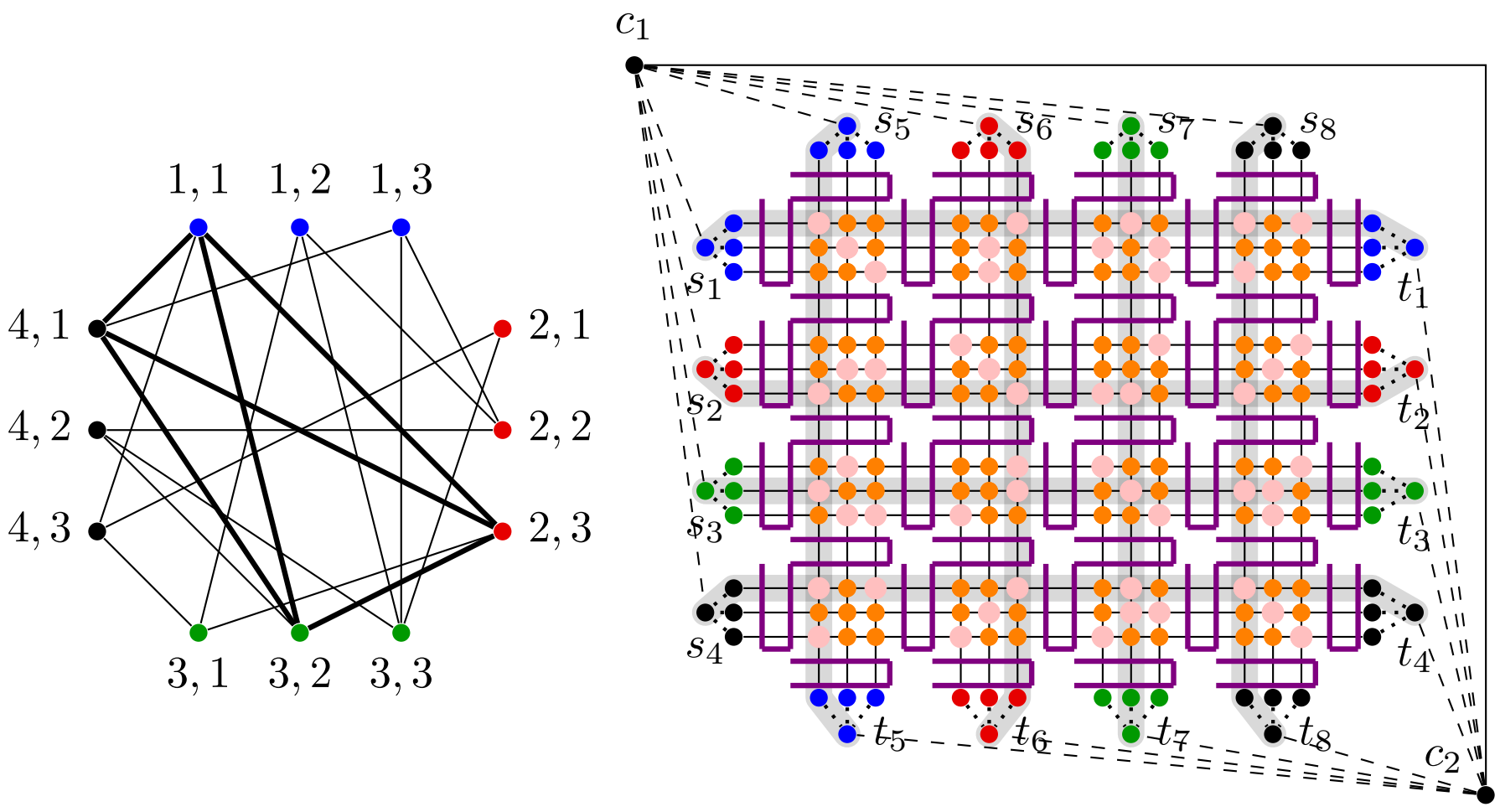%
    We start with a ``horizontal'' and a ``vertical'' path for each node in~$G$.
    We call these paths the canonical paths and they will form a grid-like structure in~$G'$ where all paths corresponding to nodes of the same color appear consecutively.
    We call paths corresponding to nodes of the same color bundles.
    Note that there is exactly one horizontal and one vertical bundle for each color.
    Wherever a vertical and a horizontal path intersect, we add either a pass-through gadget or an intersection gadget as follows.
    Say the vertical path corresponds to a node~$u$ and the horizontal path corresponds to a node~$v$.
    If~$u$ and~$v$ have different colors, then we place a pass-through gadget if~$\{u,v\} \in E$ and an intersection gadget if~$\{u,v\}$ is not a link in~$G$.
    If~$u$ and~$v$ have the same color, then we place a pass-through gadget if and only if~$u = v$.
    Between any two consecutive gadgets on any path, we add two intermediate vertices.
    Next, we add blocking gadgets.
    We only describe the blocking gadgets for horizontal paths, but the procedure is exactly the same for vertical paths and depicted in \cref{fig:hardness-example}.
    For each horizontal bundle, we add a blocking gadget between two consecutive vertical bundles with~$\nu - 1$ triples~$(s_j,t_j,1)$.
    We also add blocking gadgets before the first and after the last bundle.
    We proceed by placing~$2k$ triples, one horizontal and one vertical terminal pair for each color~$i \in [k]$.
    Let~$(s_i,t_i,1)$ be the horizontal triple for color~$i$ and~$(s_{k+i},t_{k+i},1)$ be the vertical triple for color~$i$.
    We call all of these triples the \emph{important} triples.
    We add a path of length~$25n$ between each~$s_i$ and the first node of the horizontal path corresponding to each node of color~$i$.
    We also add such paths between the last node of each of these paths and~$t_i$ and do the same for~$s_{k+i}, t_{k+i}$, and vertical paths.
    Finally, we add two new nodes~$c_1$ and~$c_2$.
    We connect~$c_1$ to all nodes~$s_i$ and~$c_2$ to all nodes~$t_i$ with~$i \in [2k]$ via a shortcut gadget.
    We add the link~$\{c_1,c_2\}$ to conclude the construction.

    Note that the constructed graph is planar as each gadget is planar and all gadgets can be arranged in a planar way as shown in \cref{fig:hardness-example}. Moreover, we set~${\ell = 1}$, ${r_i = 1}$ for each triple~$(s_i,t_i,r_i)$, and~$w(e) = 1$ for each link~$e$.
    Since the instance can be constructed in polynomial time, it only remains to show that the constructed instance is equivalent to the original instance of \textsc{Multicolored Clique}.
    Before doing so, we note that the distance between~$s_i$ and~$t_i$ for any~${i \in [2k]}$ is~$50n+9$ as two shortcut gadgets have length~$25n+4$ each.
    These shortest paths are unique and cannot be part of a solution as this would require additional flow through a shortcut gadget, which is infeasible as shown above.
    Moreover, the shortest paths between~$s_i$ and~$t_i$ that do not pass through~$c_1$ or~$c_2$ have~$L = 50n + k \cdot (5 \nu) + (k+1) (10\nu) + 1$.
    Each vertical/horizontal path corresponding to a node of color~$i$ (or~$i - k$) has this length as it contains two paths of length~$25n$ each (the dotted lines in \cref{fig:hardness-example}), $k$ bundles with~$\nu$ paths of the other orientation each and for each such path it passes through an intersection or pass-through gadget which increases the path by five each (two links in the gadget and three links to the beginning of the next gadget).
    Moreover, it passes through~$k+1$ blocking gadgets which add a length of~$10\nu$ each.
    The final plus one is to address the fact that if a shortest path between two nodes~$u$ and~$v$ contains~$x$ internal nodes, then the distance between~$u$ and~$v$ is~$x+1$.
    Note that~$L < 75n$ and hence any path that uses at least three paths of length~$25n$ is longer than any canonical path.

    We proceed to show that the constructed instance of \fste{} is equivalent to the original instance of \textsc{Multicolored Clique}.
    To this end, first assume that the original instance is a yes-instance, that is, there is a set~$K$ of nodes containing exactly one node of each color that form a clique.
    For each important triple, we route a shortest path along the canonical path representing the node in~$K$ of the respective color by choosing an appropriate waypoint on the middle of that path.
    Note that all of these paths can be routed without using any link in multiple paths as all of these paths meet in pass-through gadgets by construction.
    Moreover, as we only use one path in each blocking gadget, we can route paths for all triples with endpoints in blocking gadgets without leaving the blocking gadget by choosing appropriate waypoints (as shown in \cref{fig:gadgets}).
    Finally, for each triple with endpoints in a shortcut gadget, we use the single link between the respective terminal nodes.
    Note that all of the described paths are pairwise link-disjoint and hence we can send one unit of flow over each of them to conclude that the constructed instance of \fste{} is a yes-instance.

    Now assume that the constructed instance of \fste{} is a yes-instance and let~$w_i$ be the waypoint chosen for each triple~$(s_i,t_i,1)$ in a solution.
    As shown above, no important triple can have any of their flow be routed through a shortcut gadget.
    Note that if the shortcut gadgets are not used, then for each triple~$(s_i,t_i,1)$ with~$i \in [k]$ one unit of flow is sent through~$k+1$ blocking gadgets placed on horizontal paths and the same holds for all triples~$(s_{k+i},t_{k+i},1)$ with~$i \in [k]$ and blocking gadgets placed on vertical paths.
    Note that we showed above that each blocking gadget only allows for one unit of flow to be sent through it when ignoring the flow that comes from triples with endpoints inside the gadget.
    Since we place exactly~$k+1$ blocking gadgets for each bundle of canonical paths, this implies that no flow is routed through a blocking gadget placed on a vertical path for a triple~$(s_i,t_i,1)$ with~$i \in [k]$.
    The same holds for blocking gadgets placed on horizontal paths and triples~$(s_{k+i},t_{k+i},1)$ with~$i \in [k]$.
    This implies that each waypoint~$w_i$ for~$i \in [2k]$ is chosen from a canonical path since all other vertices are too close to one of their respective endpoints and a shortest path to the other endpoint returns to the first endpoint and then uses the shortcut gadgets.
    For each~$i \in [2k]$, let~$v_i$ be the node corresponding to the canonical path on which~$w_i$ lies.
    First, note that by construction, there is a unique shortest path from~$s_i$ to~$w_i$ and from~$w_i$ to~$t_i$.
    This path is the canonical path representing~$v_i$.
    Note that this implies that the canonical paths representing~$v_i$ and~$v_{k+j}$ for~$i,j \in [k]$ have to meet in a pass-through gadget as two units of flow are routed through each of these gadgets and without violating~$\ell = 1$, this is only possible in a pass-through gadget.
    This implies (i) $v_i = v_{k+i}$ for each~$i \in [k]$ and~$\{v_i,v_{k+j}\} \in E$ for each~$i \neq j \in [k]$.
    Hence, the nodes~$v_1,v_2,\ldots,v_k$ form a clique in the original graph~$G$.
    Moreover, since all of these nodes come from different bundles, they have pairwise different colors.
    Thus, they form a multicolored clique and the original instance of \textsc{Multicolored Clique} is a yes-instance.
\end{proof}

We next show that the variable-length variant is also NP-hard. The hardness even holds for planar graphs and unit demands.

\begin{theorem}
    \vte{} is NP-hard even if the input contains an undirected planar graph of maximum degree~$3$, $\ell = 1$, and~$r_i=1$ for all~${i \in [k]}$.
\end{theorem}

\begin{proof}
    We present a reduction from \textsc{Planar 2P1N-3-Sat} similar to the one by \citet{MP93}.
    In \textsc{Planar 2P1N-3-SAT}, we are given a formula~$\Phi$ in CNF where each clause contains two or three literals and each variable appears at most twice positively and exactly once negated in a clause in $\Phi$.
    Moreover, we are given a planar embedding of the (bipartite and planar) graph~$H$ that contains a node~$v_i$ for each variable~$x_i$ and a node~$u_j$ for each clause~$C_j$ and a link~$\{u_j,v_i\}$ if and only if~$C_j$ contains~$x_i$ or~$\neg x_i$.
    First, we take the planar embedding of~$H$ and for each link~$e$ in the embedding representing that a variable~$x$ appears in a clause~$A$, we add two nodes~$\alpha(x,A)$ and~$\beta(x,A)$ and a link between them such that~$\alpha(x,A)$ is directly left of~$\beta(x,A)$ in the planar embedding and the link (embedded as a straight line) crosses exactly the link~$e$ in the embedding of~$H$.
    Next, we remove the link~$e$ and replace the nodes in~$H$ by the gadgets in \cref{fig:clause2,fig:clause3,fig:variable} for variables, clauses with two literals, and clauses with three literals, respectively.
    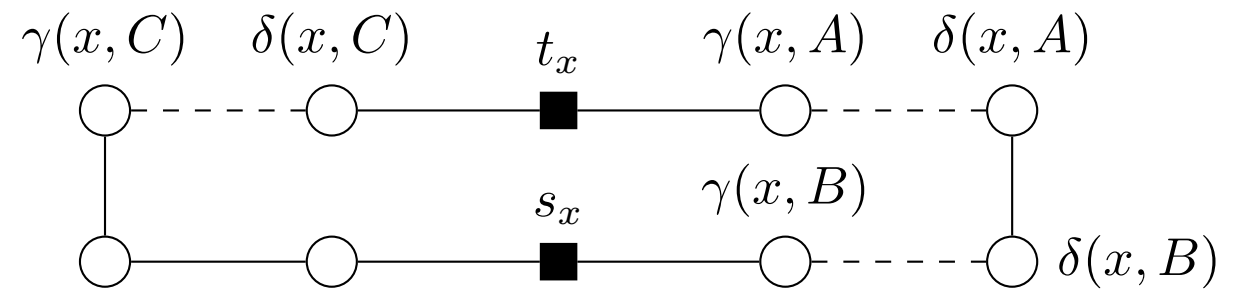
\begin{figure}[t]
        \centering
        \begin{tikzpicture}[yscale=.5]
            \node[rectangle,draw,fill=black,label=$s_x$] (u) at(0,0) {};
            \node[rectangle,fill=black,draw,label=$t_x$] (v) at(0,2) {};
            \node[circle,draw] (d1) at(-3,0) {};
            \node[circle,draw] (d2) at(-1.5,0) {};
            \node[circle,draw,label={$\gamma(x,B)$}] (b1) at(1.5,0) {};
            \node[circle,draw,label=right:{$\delta(x,B)$}] (b2) at(3,0) {};
            \node[circle,draw,label={$\gamma(x,C)$}] (c1) at(-3,2) {};
            \node[circle,draw,label={$\delta(x,C)$}] (c2) at(-1.5,2) {};
            \node[circle,draw,label={$\gamma(x,A)$}] (a1) at(1.5,2) {};
            \node[circle,draw,label={$\delta(x,A)$}] (a2) at(3,2) {};

            \draw (d2) to (u);
            \draw (d1) to (d2);
            \draw (b1) to (u);
            \draw (c1) to (d1);
            \draw (a2) to (b2);
            \draw (v) to (c2);
            \draw (a1) to (v);
            \draw[dashed] (a1) to (a2);
            \draw[dashed] (b1) to (b2);
            \draw[dashed] (c1) to (c2);
        \end{tikzpicture}
        \caption{The variable gadget for a variable~$x$ appearing positively in clauses~$A$ and~$B$ and negated in~$C$. The rectangular nodes represent terminal nodes and the dashed links are the links introduced before. The $\gamma$- and~$\delta$-nodes are the previously introduced~$\alpha$- and~$\beta$-nodes. We choose which one is which as to preserve planarity. If the variable~$x$ appears only once positively, then we replace~$\gamma(x,B)$ and~$\delta(x,B)$ and the dashed link between them by two new nodes with a regular link between them.}
        \label{fig:variable}
    \end{figure}
    \begin{figure}[t]
        \centering
        \begin{tikzpicture}[yscale=.5]
            \node[rectangle,draw,fill=black,label=left:$s_A$] (u) at(0,0) {};
            \node[rectangle,draw,fill=black,label=right:$t_A$] (v) at(2,0) {};
            \node[circle,draw,label={$\gamma(x,A)$}] (x1) at(0,1) {};
            \node[circle,draw,label={$\delta(x,A)$}] (x2) at(2,1) {};
            \node[circle,draw,label=left:{$\gamma(y,A)$}] (y1) at(0,-1) {};
            \node[circle,draw,label=right:{$\delta(y,A)$}] (y2) at(2,-1) {};

            \draw (u) to (x1);
            \draw (x2) to (v);
            \draw (u) to (y1);
            \draw (y2) to (v);
            \draw[dashed] (x1) to (x2);
            \draw[dashed] (y1) to (y2);
        \end{tikzpicture}
        \caption{The clause gadget for a clause~$A$ with two variables~$x$ and~$y$.}
        \label{fig:clause2}
    \end{figure}
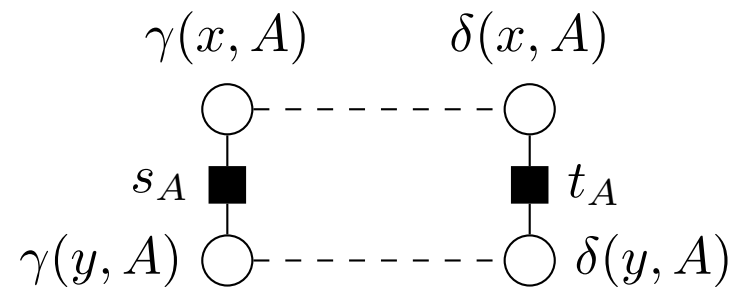
    
    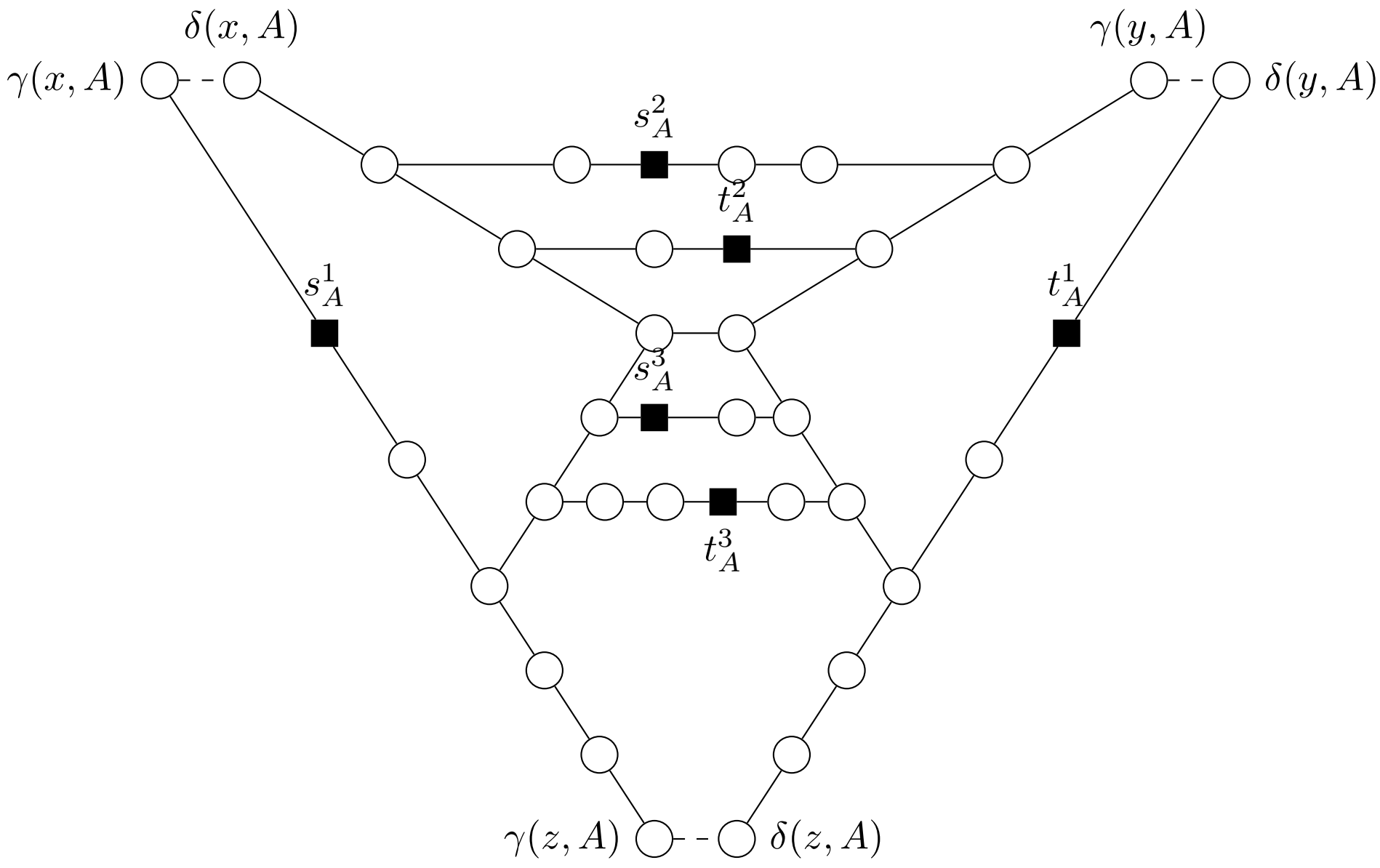
\begin{figure*}[t]
        \centering
        \begin{tikzpicture}[xscale=.75,yscale=1.15]
            \node[circle,draw,label=left:{$\gamma(x,A)$}] (x1) at(-6,6) {};
            \node[circle,draw,label={$\delta(x,A)$}] (x2) at(-5,6) {};
            \node[circle,draw,label={$\gamma(y,A)$}] (y1) at(6,6) {};
            \node[circle,draw,label=right:{$\delta(y,A)$}] (y2) at(7,6) {};
            \node[circle,draw,label=left:{$\gamma(z,A)$}] (z1) at(0,0) {};
            \node[circle,draw,label=right:{$\delta(z,A)$}] (z2) at(1,0) {};
            \node[circle,draw] (w1) at(0,4) {};
            \node[circle,draw] (w2) at(1,4) {};
            \node[rectangle,draw,fill=black,label=$s^1_A$] (v1) at(-4,4) {};
            \node[circle,draw] (v2) at(-3,3) {};
            \node[circle,draw] (v3) at(-2,2) {};
            \node[circle,draw] (v4) at(-1.333,1.333) {};
            \node[circle,draw] (v5) at(-.666,.666) {};
            \node[rectangle,draw,fill=black,label=$t^1_A$] (u1) at(5,4) {};
            \node[circle,draw] (u2) at(4,3) {};
            \node[circle,draw] (u3) at(3,2) {};
            \node[circle,draw] (u4) at(2.333,1.333) {};
            \node[circle,draw] (u5) at(1.666,.666) {};
            \node[circle,draw] (r1) at(-.666,3.333) {};
            \node[circle,draw] (r2) at(-1.333,2.666) {};
            \node[circle,draw] (s1) at(1.666,3.333) {};
            \node[circle,draw] (s2) at(2.333,2.666) {};
            \node[circle,draw] (p1) at(-3.333,5.333) {};
            \node[circle,draw] (p2) at(-1.666,4.666) {};
            \node[circle,draw] (q1) at(4.333,5.333) {};
            \node[circle,draw] (q2) at(2.666,4.666) {};
            \node[circle,draw] (a1) at(-1,5.333) {};
            \node[rectangle,draw,fill=black,label=$s^2_A$] (a2) at(0,5.333) {};
            \node[circle,draw] (a3) at(1,5.333) {};
            \node[circle,draw] (a4) at(2,5.333) {};
            \node[circle,draw] (b1) at(0,4.666) {};
            \node[rectangle,draw,fill=black,label=$t^2_A$] (b2) at(1,4.666) {};
            \node[rectangle,draw,fill=black,label=$s^3_A$] (c1) at(0,3.333) {};
            \node[circle,draw] (c2) at(1,3.333) {};
            \node[circle,draw] (d1) at(-.6,2.666) {};
            \node[circle,draw] (d2) at(.133,2.666) {};
            \node[rectangle,draw,fill=black,label=below:$t^3_A$] (d3) at(.833,2.666) {};
            \node[circle,draw] (d4) at(1.6,2.666) {};
            
            \foreach \x/\y in {x1/v1,v1/v2,v2/v3,v3/v4,v4/v5,v5/z1,z2/u5,u5/u4,u4/u3,u3/u2,u2/u1,u1/y2,x2/p1,p1/p2,p2/w1,w1/w2,w2/q2,q2/q1,q1/y1,w1/r1,r1/r2,r2/v3,w2/s1,s1/s2,s2/u3,p1/a1,a1/a2,a2/a3,a3/a4,a4/q1,p2/b1,b1/b2,b2/q2,r1/c1,c1/c2,c2/s1,r2/d1,d1/d2,d2/d3,d3/d4,d4/s2}
            {
            \draw (\x) to (\y);
            }

            \foreach \x/\y in {x1/x2,y1/y2,z1/z2}
            {
            \draw[dashed] (\x) to (\y);
            }
        \end{tikzpicture}
        \caption{The clause gadget for a clause~$A$ containing three variables~$x,y$ and~$z$.}
        \label{fig:clause3}
    \end{figure*}
    
    We next show that all gadgets can be rotated and/or mirrored to ensure that they can be embedded in a planar way given the already placed (dashed) links.
    We start with variables. To this end, let~$v_i$ be a node in~$H$ corresponding to a variable~$x_i$ in~$\Phi$.
    Let~$C$ be the clause in which~$x_i$ appears negated and let~$A$ and~$B$ be the clauses in which~$x_i$ appears positively (or ignore~$B$ if~$x_i$ only appears positively once).
    For the sake of presentation, we assume that the node representing clause~$C$ is to the left of~$v_i$ and the nodes representing~$A$ and~$B$ are to the right with the one representing~$A$ being higher in the planar embedding of~$H$.
    If this is not the case, then we can rotate the gadget appropriately and/or swap the nodes~$\gamma(x,A)$ with~$\gamma(x,B)$ and~$\delta(x,A)$ with~$\delta(x,B)$.
    Similarly, for a clause gadget for a clause containing two literals, we can rotate the gadget to ensure that the gadget can be added to the current graph without introducing link crossings.
    For clauses containing three literals~$x,y$, and~$z$, we can do the same if the variables appear in clockwise fashion. If they appear in counter-clockwise fashion, then we can mirror the gadget and then rotate it appropriately.
    Note that any sequence of three variables is ordered either clockwise or counter clockwise.
    Thus, the entire graph can be embedded in the plane without link crossings.
    To conclude the construction, we set~$T$ to be the set of all triples~$(s,t,1)$ for which~$s$ and~$t$ are the terminal pairs shown in \cref{fig:variable,fig:clause2,fig:clause3}.
    
    We next prove that the input formula~$\Phi$ is satisfiable if and only if the produced instance of \vte{} is a yes-instance.
    Since the reduction can be computed in polynomial time, \textsc{Planar 2P1N-3-Sat} is NP-hard, and the resulting graph has maximum degree at most three, it follows that \vte{} is NP-hard even if the input graph is planar and has maximum degree three.

    First, assume that~$\Phi$ is satisfiable and let a satisfying assignment~$a$ be given.
    We next show how~$a$ can be used to show that the produced instance of \vte{} is a yes-instance.
    To this end, we set the length of links between~$\alpha(x,A)$ and~$\beta(x,A)$ to 1 for all pairs of a variable~$x$ and a clause~$A$ where~$A$ contains~$x$.
    For each variable~$x$, if~$a$ assigns~$x$ to true, then we set the length of all links in the left half of \cref{fig:variable} to 1 and of all links in the right half to 2 (ignoring the dashed links).
    If~$a$ assigns~$x$ to false, then we set the length of all links in the right half to 1 and of all links in the left half (again ignoring dashed links) to~2.

    For each clause~$A$, let~$x_A$ be a variable such that the assignment of~$x_A$ with respect to~$a$ satisfies~$A$.
    If multiple such variables exist, then we choose one arbitrarily.
    If~$A$ contains two variables and~$x_A$ corresponds to~$x$ in \cref{fig:clause2}, then we set the length of all links in the upper half of \cref{fig:clause2} to~1 and the length of both non-dashed links in the lower half to~$2$.
    If~$x_A$ corresponds to~$y$ in \cref{fig:clause2}, then we set the length of all links in the lower half to~1 and the length of both non-dashed links in the upper half to~$2$.
    Finally, if~$A$ contains three literals, then there is a unique shortest path~$P_1$ (in the unweighted graph) between the first terminal pair in \cref{fig:clause3} in such a way that there exist shortest paths~$P_2$ and~$P_3$ (again in the unweighted graph) between the second and third terminal pair, respectively, such that~$P_1$ passes through~$\alpha(x_A,A)$ and~$\beta(x_A,A)$, all three paths are pairwise node disjoint and neither~$P_2$ nor~$P_3$ use any of the~$\alpha$- or~$\beta$-nodes.
    We set the length of all links in~$P_1,P_2,$ and~$P_3$ to~$1$ and all other links (ignoring dashed links) to~$2$.
    It is easy to see that all shortest paths between any terminal pair~$(s_i,t_i)$ are unique, stay within a clause gadget or within a variable gadget, and only use links that we assigned length~$1$.
    By construction, these only overlap in the nodes~$\alpha(x,A)$ and~$\beta(x,A)$.
    Since~$a$ is a satisfying assignment, whenever the clause gadget for~$A$ uses these two nodes, then the variable gadget for~$x$ does not use them.
    Thus, all terminal paths are pairwise node-disjoint (and therefore also link-disjoint) and thus we constructed a solution for~$\ell=1$.
    We pick one waypoint on each of these paths arbitrarily.
    
    For the other direction, assume that the constructed instance of \vte{} is a yes-instance.
    In particular, there are pairwise edge-disjoint paths between all terminal pairs.
    Since the maximum degree is three and terminals have degree 2, these paths are also vertex-disjoint.
    Since all links in the clause gadget depicted in \cref{fig:clause2} are incident to terminal vertices (or are dashed links), the only way for a solution of a variable gadget to leave the variable gadget is to pass through at least two gadgets in \cref{fig:clause3}.
    However, routing any additional path through the gadget in \cref{fig:clause3} is impossible.
    This was shown by \citet{MP93} and can be summarized in the following observation.
    If an additional path is routed through the gadget, then this path uses at least one endpoint of two of the dashed links.
    If all three dashed links are unavailable, then it is impossible to route paths between all three terminal pairs in the gadget and using exactly one of the dashed links requires to use both central nodes or the dashed link~$\{\gamma(z,A),\delta(z,A)\}$.
    In the former case, no additional path can be routed through the gadget and in the latter case, only a path from~$\delta(x,A)$ to~$\gamma(y,A)$ can be routed.
    This path, however, makes it impossible to connect the terminal pair~$(s_A^2,t_A^2)$.
    Thus, solution path between terminal vertices in the variable gadget is completely contained in the variable gadget.
    Let now~$b$ be an assignment to the variables of the original formula~$\Phi$ that assigns a variable to true if the solution path between the terminal vertices in this variable's gadget uses the left half (in \cref{fig:variable}).
    We claim that~$b$ satisfies~$\Phi$.
    Assume towards a contradiction that there is some clause~$A$ in $\Phi$ that is not satisfied by~$b$.
    This means that none of the variables appearing in~$A$ satisfy the clause under assignment~$b$.
    By construction, this means that solution path between the terminal vertices in all variable gadgets for a variable~$x$ in~$A$ uses the respective pair of~$\alpha(x,A)$ and~$\beta(x,A)$ vertices.
    Hence, the solution paths between the terminal vertices in the clause gadget for~$A$ (depicted in either \cref{fig:clause2} or \cref{fig:clause3}) cannot use the~$\alpha$- and~$\beta$-vertices ($\gamma$ and~$\delta$ in the pictures).
    Note that without these vertices, there is only a single path between the first terminal pair in \cref{fig:clause3} that avoids the other terminals.
    Moreover, no such path exists between the terminal pair in \cref{fig:clause2}.
    However, for the unique path between the first terminal pair in \cref{fig:clause3} that avoids all~$\alpha$- and~$\beta$-vertices, there is no vertex-disjoint path between the third terminal pair $(s_A^3,t_A^3)$.
    Hence, the assumption that~$b$ is not a satisfying assignment contradicts the fact that we assumed the constructed instance to be a yes-instance.
    This concludes the proof.
\end{proof}

We conclude by showing that \fste{} can be solved by the relatively simple mixed integer linear program presented in \cref{fig:modILP}.
\begin{proposition}
    \label{prop:ILP}
    \fste{} can be encoded in a mixed integer linear program with~$nk$ integer variables and $O(k\cdot (n+m))$ (continuous) variables and constraints.
\end{proposition}

\begin{proof}
    For each~$i \in [k]$ and each node~$v \in V$, we add a variable~$w^i_v \in \{0,1\}$.
    This variable will be set to~$1$ if~$v$ is the waypoint for the triple~$(s_i,t_i,r_i)$ in a solution.
    We precompute all distances in the graph in~$O(n^3)$ time using the Floyd–Warshall algorithm \cite{Floyd62}.
    We denote by~$\dist(u,v)$ the distance between~$u$ and~$v$ in~$G$. 
    Note that a link~$\{u,v\}$ can only be used on a shortest path between~$s_i$ and~$w_i$ (or~$t_i$ and~$w_i$) if~$\dist(s_i,u) = \dist(s_i,v) \pm w(\{u,v\})$.
    Moreover, it can only be used in one direction (from the node closer to~$s_i$ to the node further away from~$s_i$).
    For this reason, we can define a directed graph~$(V,A_{s_i})$ where~$A_{s_i}$ contains all directed links whose undirected counterpart is contained in~$E$ and where the link points away from~$s_i$.
    That is, we set~$A_{s_i} = \{(u,v) \mid \{u,v\}\in E \land d(s,v) = d(s,u) + w(\{u,v\})\}$.
    If~$G$ is already directed, this is just a subgraph of~$G$.
    We do the same for each~$t_i$ and call the resulting link set~$A_{t_i}$.
    That is, $A_{t_i} = \{(u,v) \mid \{u,v\}\in E \land d(v,t) = d(u,t) + w(\{u,v\})\}$.
    Note that this reverses the arcs in case~$G$ is directed.
    
    The constraints of our mixed integer linear program for undirected graphs are given in~\cref{fig:modILP}.
    It uses variables~$w_v^i$ for each~$i\in [k]$ and~$v \in V$, $x_a^i$ for each~$i \in [k]$, and~$a \in A_{s_i}$ and~$y_a^i$ for each~$i \in [k]$ and~$a \in A_{t_i}$.
    Each of these variables can be assigned a rational number between~$0$ and~$r_i$.
    Finally, we have one variable~$\ell \geq 0$ that restricts the maximum flow sent over any link in~$G$ in a solution.
    The goal is to compute a solution which minimizes~$\ell$.
    We mention that constraint (2) includes some terms that are not valid variables as not all edges have to be included in~$A_{s_i}$ and/or~$A_{t_i}$.
    Those are of course not included in the sum, but we kept this presentation for the sake of readability.
    \begin{figure*}[t]
    \hfill\scalebox{.975}{\parbox{\linewidth}{
    \begin{align}
         &\sum_{v \in V} w_v^i = 1 & \text{ for all } i\in [k]\\
         &\sum_{i \in [k]} x_{(u,v)}^{i} + x_{(v,u)}^{i} + y_{(u,v)}^{i} + y_{(v,u)}^{i} \leq \ell & \forall \{u,v\} \in E\\
         &\sum_{(s_i,v) \in A_{s_i}} x_{(s_i,v)}^{i} = r_i - r_i \cdot w_{s_i}^i & \forall i\in [k]\\
         &\sum_{(u,v) \in A_{s_i}} x_{(u,v)}^{i} = \sum_{(v,u) \in A_{s_i}} x_{(v,u)}^{i} + r_i \cdot w_v^i & \forall v \in V \setminus \{s_i\}, i\in [k]\\
         &\sum_{(t_i,v) \in A_{t_i}} y_{(t_i,v)}^{i} = r_i  - r_i \cdot w_{t_i}^i & \forall i\in [k]\\
         &\sum_{(u,v) \in A_{t_i}} y_{(u,v)}^{i} = \sum_{(v,u) \in A_{t_i}} y_{(v,u)}^{i} + r_i \cdot w_v^i & \forall v \in V \setminus \{t_i\}, i\in [k]
    \end{align}
    }}
    \caption{A mixed integer linear program for \fste{} for undirected graphs. For directed graphs, replace Constraint (2) by $\sum_{i \in [k]} x^i_{(u,v)}+y^i_{(v,u)} \leq \ell$ for all~$(u,v) \in E$.}
    \label{fig:modILP}
    \end{figure*}

    The basic idea is that constraint~(1) guarantees that precisely one waypoint~$w_i$ is picked for each triple~$(s_i,t_i,r_i)$.
    The flow between~$s_i$ and~$w_i$ (between~$t_i$ and~$w_i$, respectively) is then captured by the assignment of~$x$ (respectively $y$) variables as follows.
    If we assign~$x_{u,v}^{i}$ some positive value~$p$, then this corresponds to sending~$p$ units of flow from~$u$ to~$v$.
    We do the same for~$y_{u,v}^i$.
    Here, we use the arc in the opposite direction and produce a flow ``from~$t_i$ to~$w_i$''.
    Constraints (3) and (5) then ensure that flow~$r_i$ leaves~$s_i/t_i$ unless that is the chosen waypoint.
    Constraints (4) and (6) are the usual flow-preservation constraints.
\end{proof}

We mention that a similar mixed integer linear program can also be used to solve the variant where we need to select a single path between each terminal pair.
Simply constrain the~$x$ and $y$ variables to be assigned either~0 or~1, remove~$r_i$ in all constraints, and add it to constraint~(2) instead.

\section{Conclusion}\label{sec:conclusion}

We charted a refined landscape of the complexity of traffic engineering in segment routing networks which recently received much interest in industry. In particular, we studied the practically relevant case of planar graphs and showed that two very basic models remain NP-hard: when link weights are fixed and the operator may optimize the splitting of traffic among shortest paths and when link weights and waypoints may be chosen and every segment must follow a unique shortest path. Our results also show that the computational difficulty is fundamentally combinatorial and not caused by the numbers involved. At the same time, our compact mixed integer linear formulation for the fixed-weight splittable setting provides a direct approach for computing exact solutions and a baseline for evaluating other approaches.

\section*{Acknowledgements}
This research is supported by the German Research Foundation (DFG), SPP 2378: ReNO-2 (project number 511099228).

\bibliographystyle{plainnat}
\bibliography{bib}

\end{document}